\documentstyle[12pt]{article}

\newcommand{\EQ}{\begin{equation}}
\newcommand{\EN}{\end{equation}}

\renewcommand{\thefootnote}{\fnsymbol{footnote}}
\def\aprle{\buildrel < \over {_{\sim}}}
\def\aprge{\buildrel > \over {_{\sim}}}
\begin{document}
\topmargin 0pt
\oddsidemargin=-0.4truecm
\evensidemargin=-0.4truecm
\newpage
\setcounter{page}{0}
\rightline{IC/94/42}

\begin{center}
{\normalsize International Atomic Energy Agency\\
and\\
United Nations Educational, Scientific and Cultural Organization\\}
\vskip 1mm
{
INTERNATIONAL CENTRE FOR THEORETICAL PHYSICS}

\vskip 10mm

{\bf \Large
Neutrinos with Zee-Mass Matrix
in Vacuum and Matter
}

\vskip 10mm
{Alexei Yu. Smirnov
\footnote{On leave from Institute for Nuclear Research, Russian Academy of
Sciences, 117312 Moscow, Russia.\\
\phantom{okji}e-mail: smirnov@ictp.trieste.it} and Zhijian Tao}\\
{International Centre for Theoretical Physics, I-34100 Trieste, Italy}\\

\vskip 9mm
ABSTRACT
\end{center}
Neutrino mass matrix generated by the Zee (radiative) mechanism has
zero (in general, small) diagonal elements and a natural hierarchy of the
nondiagonal elements. It can be considered as an alternative
(with strong predictive power) to the
matrices generated by the see-saw mechanism.
The propagation in medium of the neutrinos with the Zee-mass matrix
is studied. The  flavor neutrino transitions are
described  analytically. In the physically interesting
cases the probabilities of transitions as functions of neutrino
energy can be represented
as two-neutrino probabilities modulated by the effect of vacuum
oscillations related to the small mass splitting.
Possible applications of the results to the
solar, supernova, atmospheric and relic neutrinos are discussed.
A set of the predictions is found
which could allow to identify the  Zee-mass matrix and therefore the
corresponding mechanism of mass generation.

\vskip 1truecm
\begin{center}
MIRAMARE -- TRIESTE\\

\vskip 1mm

March 1994
\end{center}
\vspace{1cm}
\vspace{.5cm}
\renewcommand{\thefootnote}{\arabic{footnote}}
\setcounter{footnote}{0}

\newpage

\section{ Introduction}

There are a number of mechanisms of the neutrino mass generation
which  relate  naturally the
smallness of the  masses to the neutrality of neutrinos. Those include
the see-saw mechanism \cite{ss},\cite{goran} different  radiative
mechanisms \cite{zee}, \cite{barb}, \cite{rpar}, \cite{bh}, \cite{bzf},
\cite{babu}, the tree level mass generation by Higgs triplet having small
induced VEV \cite{trip}, mass
generation by nonrenormalizable Planck scale interaction
\cite{planck}. In general several  mechanisms
contribute to the neutrino masses simultaneously, moreover their
contributions can be
comparable or one of them dominates.  Different mechanisms
imply different symmetries and particle contents of the theory.

There are some  indications of the  existence of tiny
neutrino masses and lepton mixing related
to the solar, atmospheric and relic neutrinos.
The suggested mechanisms can rather
easily generate the neutrino masses (mass squared differences)
and mixings in some or even all regions of these
``positive indications".
What can be learned about the origin of  neutrino masses
and mixings from the existing, and mainly, from future
experiments with solar, atmospheric,  supernova neutrinos as well as
from the accelerator experiments? Can one conclude something on the
mechanism of the neutrino mass generation?
In principle,  different mechanisms result in different structures of
mass matrices and consequently in different features of the neutrino
propagation in vacuum and matter. For the identification of the  mechanism
it is crucial to study the effects of mixing of all three
neutrinos. The propagation of three neutrinos with strong hierarchy of
masses and mixings  typical for simplest versions of
the see-saw mechanism has been widely discussed before\cite{three}.

In this paper we will study the properties of propagation
in medium of the  neutrinos $\nu_f^T\equiv(\nu_e, \nu_{\mu}, \nu_{\tau})$
having  the Majorana mass matrix:
\begin{equation}
M=\left|\begin{array}{lll}
0&m_{e\mu}&m_{e\tau}\\
m_{e\mu}&0&m_{\mu\tau}\\
m_{e\tau}&m_{\mu\tau}&0\\
\end{array}\right|,~~~~
m_{e \mu} \ll m_{\mu \tau}, m_{e \tau} ~.
\end{equation}
Its crucial features are: zero (in general small)
diagonal elements and strong hierarchy of the nondiagonal elements.
Such a matrix appears in a class of models with radiative mass generation.
The first and the simplest version  which naturally results
in structure (1) had been suggested long time ago by Zee \cite{zee}
and we will call matrix (1)  Zee-mass matrix.

The Zee mechanism implies the
existence of charged scalar field, $S$, singlet of the $SU(2)$
and two doublets of Higgs bosons  $\Phi_u$, $\Phi_d$.
The interactions responsible for the neutrino mass generation are
$$
\sum_{\alpha \beta} f_{\alpha \beta}
\Psi_{\alpha L}^TC\tau_2\Psi_{\beta L}S^{\dagger} +
\sum_{\alpha} \frac{m_{\alpha}}{v_d} \bar{\Psi}_{\alpha L} l_{\alpha R}
\Phi_d + M_{ud}\Phi_u\tau_2\Phi_d S^{\dagger} + h.c.,
$$
($\alpha, \beta = e, \mu, \tau)$, where $\Psi_{\alpha L}$ is the lepton
doublet,  $l_{\alpha R}$ is the right handed component of the charge
lepton,  $m_{\alpha}$ is its mass,  and
$M_{ud}$ is the mass parameter.
As a consequence of the gauge symmetry the couplings
of the singlet $S$ to lepton doublets
are antisymmetric: $f_{\alpha \beta} = -f_{\beta\alpha}$.
Only one doublet, $\Phi_d$, with vacuum expectation
$v_d$ gives  masses to the charged leptons. The indicated interactions
generate at one loop level
the elements of the mass matrix (1) \cite{zee} \cite{petcov}
\begin{equation}
m_{\alpha\beta} \simeq \frac{f_{\alpha \beta}}
{16 \pi^2}(m_{\alpha}^2 - m_{\beta}^2)
\frac{I}{v}~,\hskip 1cm
I\equiv \frac{v_u}{v_d} \cdot
\frac{M_{ud} v }{M_2^2 - M_1^2}\cdot ln \frac{M_2^2}{M_1^2}~,
\end{equation}
where $M_1$, $M_2$ are the masses of charged scalars,
$v \equiv \sqrt{v_u^2 + v_d^2}$  is the electroweak scale
and $v_u$ is the vacuum expectation of $\Phi_u$.
{}From (2) one finds the relations between the masses:
\begin{equation}
\displaystyle{\frac{m_{\mu\tau}}{m_{e\tau}}
\cong\frac{f_{\mu\tau}}{f_{e\tau}}}, \hskip 1cm
\displaystyle{\frac{m_{e\mu}}{m_{e\tau}}\cong\frac{f_{e\mu}}{f_{e\tau}}
\left(\frac{m_{\mu}}{m_{\tau}}\right)^{2}},
\end{equation}
i.e.
the element $m_{e\mu}$ turns out to be naturally suppressed in
comparison with the two other elements
thus reproducing the structure (1), unless $f_{\alpha\beta}$ have strong
inverse flavor hierarchy.

The matrix (1)  appears also in some modifications of the  Zee-mechanism.
Instead of one scalar, three charged scalars,
$S_{e\mu}, S_{e \tau}, S_{\mu \tau}$, are  introduced in \cite{bh}
that carry double lepton charges.
Moreover, the lepton number can be violated
spontaneously by the VEV of new
neutral singlet which has the coupling $\Phi_u\Phi_dS^+S^0$ \cite{bh}.
Three boson coupling can be generated in   one loop \cite{bzf},
being suppressed at tree level by a discrete symmetry.
The general properties of models which result in the matrix (1)  are (i)
the existence of one (or several) charged scalar fields singlets of $SU_2$,
(ii) the generation of masses of the charged leptons at tree level by only
one Higgs
doublet. Last feature explains simultaneously a suppression of flavor
changing transitions and can be related to a certain discrete symmetry.

As can be shown, the properties of propagation are practically the same
for a more general form of mass matrix with nonzero diagonal elements
provided  such elements do not exceed appreciably $m_{e\mu}$.
The appearance of these diagonal elements can be related,
e.g. to the violation
of the discrete symmetry in the interactions of Higgs scalars.
(The restrictions on  flavor changing neutral currents
admit $m^{diag} \sim m_{e \mu}$). They
can result also from the see-saw contributions or from the radiative
effects related to the R-parity breaking. Note that
such a matrix as well as the original Zee-matrix have  an approximate
$L_e + L_{\mu} - L_{\tau}$ lepton number conservation.
This  allows for a
large mixing of e- and $\mu$- flavors. The $\tau$ flavor turns out to be
singled out.

The properties of the Zee mass matrix in vacuum have been studied
by Wolfenstein \cite{wolf}.
Here we consider the matter effects and  confront the results
with the existing data. The paper is organized as follows. In sect.2
we summarize the properties of the Zee matrix and the
oscillations induced by this matrix in vacuum. In sect.3 the properties
of the Zee matrix in matter will be considered.
Sect.4 is devoted to the conversion induced by the Zee matrix in matter.
The application will be discussed in sect.5. Sect.6 contains the
outlook and the discussion.

\section{ Zee mass matrix in vacuum}

Instead of  $m_{ij}$ it is convenient to introduce
basic mass scale, $m_0$, $e \mu$- mixing angle,
$\theta$, and small parameter, $\epsilon$,
as\footnote{In contrast with Wolfenstein \cite{wolf}, we use
$\sin\theta\leftrightarrow\cos\theta$,~$\epsilon\rightarrow\delta$.}:\\
\begin{equation}
\begin{array}{l}
m_0\equiv\sqrt{m_{e\tau}^2+m_{\mu\tau}^2}, ~~~
\sin\theta\equiv\displaystyle{\frac{m_{e\tau}}
{m_0}} =
\frac{f_{e\tau}}{\sqrt{f_{e\tau}^2 + f_{\mu\tau}^2}} ,\\
\epsilon\equiv\displaystyle{\frac{m_{e\mu}}{m_0}} =
\left( \frac{m_{\mu}}{m_{\tau}} \right)^2
\frac{f_{e\mu}}{\sqrt{f_{e\tau}^2 + f_{\mu\tau}^2}}.
\end {array}
\end{equation}
In terms of these parameters the mass
matrix (1)  can be rewritten as
\begin{equation}
M= m_0\left|\begin{array}{lll}
0& \epsilon & \sin \theta \\
\epsilon & 0 & \cos \theta \\
\sin \theta & \cos \theta &0\\
\end{array}\right|.
\end{equation}
Let us summarize the properties of this matrix. The eigenvalues of (5)
equal to \cite{wolf}
\begin{equation}
m_1=- m_0 \epsilon\sin2\theta, ~~~
m_{3,2} =  m_0(\pm 1 - \frac{1}{2}\epsilon\sin2\theta),
\end{equation}
and the mixing matrix in vacuum which relates the flavor
and the mass eigenstates
$\nu_f=S_0 \nu$  ($\nu \equiv \nu_1, \nu_2, \nu_3$)
is
\begin{equation}
S_0 = \frac{1}{\sqrt{2}}
\left(
\begin{array}{ccc}
\sqrt{2}c&s&s\\
-\sqrt{2}s&c&c\\
0&1&-1
\end{array}
\right)+O(\epsilon),
\end{equation}
here (and everywhere below) $s\equiv\sin\theta$, $c\equiv\cos\theta$.

Since $\epsilon \ll 1$,  two states, $\nu_2$ and $\nu_3$,
are approximately degenerate, and their masses
($\sim m_0$) are much larger than the
mass of $\nu_1$. Moreover
$\nu_2$ and $\nu_3$ have opposite CP-parities thus forming
the pseudo Dirac neutrino. Their mass squared difference is
\begin{equation}
\Delta m_{23}^2=2\epsilon \sin 2\theta m_0^2 \ll m_0^2.
\end{equation}
Mixing of the electron and the muon neutrinos is
determined by  $\theta$, it becomes maximal  at  $m_{e\tau}\simeq
m_{\mu\tau}$
($f_{e\tau} = f_{\mu \tau}$). The tau
neutrino is mixed almost maximally with
the combination  $\nu_2^0 = \sin \theta \nu_e + \cos \theta \nu_{\mu}$;
the deviation from maximal mixing is proportional to
$\epsilon^2$:
\begin{equation}
1 - \sin^22\theta_{\tau 2}=\frac{1}{4} \epsilon^2 \sin2\theta .
\end{equation}

The mass squared differences for $\nu_1$ component equal
$\Delta m^2_{12} \approx \Delta m^2_{13}
\approx m_0^2$ and  the ratio of mass differences (see (6)) is
determined by $\epsilon$:
\begin{equation}
\frac{\Delta m^2_{23}}{\Delta m^2_{12}} = 2\epsilon \sin2 \theta .
\end{equation}
Consequently, the Zee-matrix has two crucial features:
it gives naturally two
different
scales for the  mass squared differences  and practically maximal mixing
between the two heaviest components.

In vacuum the propagation of neutrinos having
the Zee-mass matrix results in superposition of oscillations with
two different oscillation lengths; the ratio of lengths is determined
by (10) \cite{wolf}.
For small distances, $L \ll 4\pi E/\Delta m^2_{23}$,
the task is reduced to two neutrino oscillations
$\nu_e \leftrightarrow \nu_{\mu}$  with depth  $\sin^2 2\theta$.
The probabilities of $3\nu$-oscillations
can be immediately found from (6) (7). For example, the
averaged survival probabilities
$P(\nu_e\rightarrow\nu_e)$ and $P(\nu_{\mu}\rightarrow\nu_{\mu})$ are
\cite{wolf}
\begin{equation}\begin{array}{ll}
P(\nu_e\rightarrow\nu_e)=
c^4+\displaystyle{\frac{1}{2}s^4}, &
P(\nu_{\mu}\rightarrow\nu_{\mu})=
s^4+\displaystyle{\frac{1}{2}c^4}.
\end{array}
\end{equation}
Note that the effect of the third neutrino is reduced to the factor of 1/2
in (11) which
comes from the lost of the coherence in the maximally mixed neutrino state.

\section{
Properties of Zee mass matrix in matter: levels and mixing}

{\it Evolution equation. Effective Hamiltonian.}
In matter the evolution equation of the
neutrinos $\nu_f^T\equiv(\nu_e, \nu_{\mu}, \nu_{\tau})$
can be written as \cite{msw}
\begin{equation}
i\displaystyle{\frac{d\nu_f}{dt}=H\nu_f}, ~~~
H\simeq\displaystyle{\frac{M^2}{2E}+H_{matter}},
\end{equation}
where $H$ is the effective Hamiltonian  for ultrarelativistic neutrino,
$M^2$ is the vacuum mass matrix (5) squared, E is the energy
of neutrino and $H_{matter}$ is the matrix describing matter effect:
$
H_{matter}=diag[\sqrt{2}G_Fn_e,~~ 0,~~ 0].
$
Here $G_F$ is the Fermi constant and $n_e$ is the concentration of
electrons.
In (12) we have neglected the high order electroweak effects which lead also
to  the splitting of $\nu_{\mu}-\nu_{\tau}$ levels \cite{bot} and suggest
that the concentration of  neutrinos in medium is small
(so that neutrino - neutrino
scattering which in particular, generates the nondiagonal elements
of $H_{matter}$\cite{pant} can be neglected). Substituting
the  explicit expressions for $M^2$  and $H_{matter}$ in (12) one gets
\begin{equation}
H=\displaystyle{\frac{m_0^2}{2E}}\left|
\begin{array}{ccc}
s^2+\epsilon^2+\rho&sc&\epsilon c\\
sc&c^2+\epsilon^2 &\epsilon s\\
\epsilon c&\epsilon s&1
\end{array}
\right| ,~~~~
\rho\equiv\sqrt{2}G_Fn_e\displaystyle{\frac{2E}{m_0^2}}.
\end{equation}

Diagonal elements of $H$ determine the flavor energy levels, and as the
density changes, there are two  crossings of flavor levels. The
resonance (crossing) densities are:
\begin{equation}
\rho_{e\mu}=\cos2\theta , ~~~~\rho_{e\tau} = \cos^2 \theta,
\end{equation}
for $\nu_e-\nu_{\mu}$ and $\nu_e-\nu_{\tau}$ levels correspondently
(see fig.1). Evidently $\rho_{e\tau}>\rho_{e\mu}$; the distance between
crossing points ($\rho_{e\tau}-\rho_{e\mu}=s^2$)  is always
smaller than the width of the $e-\mu$ resonance ($\Delta\rho\sim
\tan2\theta$).
Consequently, there is a strong influence of the  resonance
related to the largest $e-\mu$-mixing on that stipulated by
$e-\tau$-mixing (fig. 1).

\vskip 3mm
\noindent
{\it Mixing matrix in matter, level crossing scheme.}
Let us introduce the neutrino eigenstates in matter, $\nu_{im}$,
and corresponding eigenvalues
$H_i$, (i = 1, 2, 3) as the eigenstates and the eigenvalues of $H$.
The eigenstates and  the flavor
states are related by mixing matrix in matter, $S(\rho)$:
$\nu_f=S(\rho)\nu_m$. The matrix $S(\rho)$ and the eigenvalues
$H_i$ are determined by the diagonalization condition
$
S(\rho)^{\dagger} H(\rho) S(\rho)=diag\{H_1,~~H_2,~~H_3\} .
$

Let us find $S(\rho)$ and $H(\rho)$. As follows from (13) in zero
order over
$\epsilon$ ($m_{e\mu}=0)$ the state $\nu_{\tau}$ decouples and
the task is reduced to two neutrino case.
The Hamiltonian $H^0 \equiv  H(\epsilon = 0)$ is diagonalized by the
rotation
\begin{equation}
S_{12}(\theta_m) = \left|\begin{array}{ccc}
\cos \theta_m& \sin \theta_m & 0\\
-\sin \theta_m&\cos \theta_m & 0\\
0 & 0 & 1
\end{array}\right|,
\end{equation}
where $\theta_m$ is the  $e \mu$ - mixing angle in matter fixed by:
\begin{equation}
\tan 2\theta_m=\displaystyle{\frac{\sin 2\theta}{\cos 2\theta-\rho}}.
\end{equation}
The eigenvalues of $H^0$ which correspond to the rotated states
$\nu_m^0 \equiv (\nu_{1m}^0, \nu_{2m}^0, \nu_{\tau})$
can be written in the units $\displaystyle{\frac{m_0^2}{2E}}$ as
\begin{equation}\begin{array}{l}
H_{2(1)}^0=\displaystyle{\frac{1}{2}}\left[1 +
\rho \pm \sqrt{(1+\rho)^2-4c^2\rho}\right],~~~~
H_{\tau} = 1.
\end{array}\end{equation}
The level splitting,
$\Delta H_{12}\equiv H_2^0 - H_1^0=\sqrt{
(\cos2 \theta - \rho)^2 + \sin ^2 2\theta}$,
is minimal in resonance (14):
$\Delta H_{12}(\rho_{e\mu})=\sin 2\theta$.

At $\rho=0$, one gets from (17): $H_1^0 = 0$, and $H_2^0 = H_{\tau} = 1$, i.e.
the levels $H_2^0$ and   $H_{\tau}$  cross at zero density
(fig.1). Due to the strong influence of $e\mu$-mixing on
$e\tau$-mixing the
resonance for $\nu_{\tau}$ is shifted from $\rho_{e\tau}$
to $\rho = 0$.

The elements of the Hamiltonian (13) proportional to  $\epsilon$
give the corrections to the above
level scheme which become important when $\rho \rightarrow 0$,
in particular they induce the mixing of $\nu_{\tau} - \nu_{2m}^0$ states.
Using (13) and the mixing matrix (15) one finds the Hamiltonian
in the basis of the states $\nu_m^0 \equiv (\nu_{1m}^0, \nu_{2m}^0,
\nu_{\tau})$. Performing then an additional rotation,
$S_{13}(\alpha)$,  in $\nu_{1m}^0, \nu_{\tau}$ space,
on the
angle $\alpha$ determined by
\begin{equation}
\tan 2 \alpha = \frac{2 \epsilon \cos(\theta + \theta_m)}{1 - H_1^0 -
\epsilon^2},
\end{equation}
$(\nu_{m}^0 \rightarrow \nu_m',
\nu_m^0
= S_{13}(\alpha)\nu_m'$, and $\nu_m' \equiv
(\nu_{1m}, \nu_{2m}^0, \nu_{\tau}') ) $
one gets the Hamiltonian which describes the conversion at small
densities
\begin{equation}
H\simeq\left|
\begin{array}{ccc}
H_1^0 - \delta & 0 & 0 \\
0 & H_2^0  & \epsilon \sin(\theta + \theta_m) \cos\alpha\\
0 & \epsilon \sin(\theta + \theta_m) \cos \alpha & 1 + \epsilon^2 + \delta
\end{array}
\right|+O(\epsilon^3).
\end{equation}
Here
$$
\delta \equiv -  (1 - H_1^0) \sin ^2\alpha +
\epsilon \cos(\theta + \theta_m) \sin 2 \alpha
$$
is of the order of $\epsilon^2$,
the values of $\theta_m$, $H_1^0$, $H_2^0$ are determined in (16 - 17).

According to (18) the angle $\alpha$ being also of the order of $\epsilon$
increases with density:
$$
\tan2 \alpha = \left\{
\begin{array}{ll}
2\epsilon \cos 2 \theta, & \rho = 0 \\
2\epsilon (c + s)^{-1}, & \rho = \cos2 \theta \\
2\epsilon /s, & \rho \gg 1
\end{array}
\right.
$$
There is no resonant enhancement of $\alpha$.

The Hamiltonian (19) is diagonalized by the rotation $S_{23}(\beta)$ in
the $(\nu_{2m}^0, \nu_{\tau}')$-space by the angle $\beta$ determined from
\begin{equation}
\tan 2 \beta \approx  \frac{2 \epsilon \sin(\theta + \theta_m) \cos \alpha}
{1 - H_2^0 + \delta} \approx
\displaystyle{\frac{2\epsilon \sin 2\theta}
{-\epsilon^2\sin^2 2\theta - \rho s^2}}
 \approx \frac{4 \epsilon}{\rho \tan \theta}~.
\end{equation}
Here we have taken into account that in the denominator $\delta$ can be
neglected everywhere except  small region  around $\rho \approx 0$
and that $H_2^0 \approx 1 + \rho s^2$
for $\epsilon^2 \ll \rho \ll 1$ as follows from (17). Obviously,
$\tan 2 \beta \ll 1$
for $\rho \gg 4 \epsilon /\tan \theta$.
If $\rho \gg 1$ one has $H_2^0 \approx \rho$ and the angle
$\beta$ is even more  strongly suppressed
$\tan 2 \beta \approx 2 \epsilon / \rho$.
At $\rho=0$ second expression in (20)  reproduces the result (9).

The total mixing matrix in matter
that diagonalizes the original Hamiltonian (13) is
$
S = S_{12}(\theta_m) \cdot S_{13} (\alpha) \cdot S_{23}(\beta).
$
The eigenstates $H_2$ and $H_3$ can be
found  as the eigenstates of $2\times2$ submatrix of (19),
$H_1 = H_1^0 - \delta$.
Two angles, $\theta_m$ and $\beta$, undergo the resonant enhancement in
different density regions.

\vskip 3mm
\noindent
{\it Two density regions.}
It is possible to divide  the whole density
region into two parts
so that in each part the three level mixing  is reduced to two level mixing.
Indeed, let us define the density
\begin{equation}
\rho_b \equiv \frac{4 \epsilon}{\tan \theta}
\end{equation}
which fixes
the width of the  $\nu_{\tau} - \nu_{2m}^0$ -
resonant layer ($\Delta \rho_R = 2 \rho_b$)
(for $\rho < \rho_b$ one has  $\sin^2 2\beta < 1/2$).
If $\rho_b \ll 1$, the regions of small and large densities can be
introduced.

(i). Large density region: $\rho \gg \rho_b$.
Here $\alpha, \beta \sim \epsilon$ and
in the lowest approximation
the mixing matrix is $S \approx  S_{12}(\theta_m)$. The state
$\nu_{\tau}$ decouples; the dynamics of
propagation is determined by change of $\theta_m$.

(ii). Small density region: $\rho \aprle \rho_b$.
The change of  mixing is determined by the angle $\beta$,
whereas two other angles  vary weakly
(even if  $\theta = 45^0$) coinciding practically
with vacuum values: $\theta_m \approx \theta$,
$\tan 2 \alpha \approx 2
\epsilon \cos 2\theta$. Consequently, in the first approximation one has:
$S = S_{12}(\theta) \cdot S_{23}(\beta)$;
$\nu_{1m}^0$ decouples and the dynamics  of level
crossing is determined by $2\times2$ submatrix of (19).

For extremely large densities, $\rho \gg 1$, : $\theta_m \approx \pi/2$,
$\beta \approx 0$ and  $\alpha \approx \epsilon / s$ so that
\begin{equation}
S_m(\rho>>1)\simeq\left|
\begin{array}{ccc}
0                  & 1 & 0\\
-1                 & 0 & \frac{\epsilon}{s}\\
\frac{\epsilon}{s} & 0 & 1
\end{array}
\right|,
\end{equation}
i.e. $\nu_e$ state decouples being composed of the eigenstate $\nu_{2m}$,
whereas $\nu_{\mu}$ and $\nu_{\tau}$ are mixed with the angle
$\alpha \sim \epsilon/s$.

\vskip 3mm
\noindent
{\it Other possibilities.}
In special case of equal $f_{\alpha \beta}$, one has
$m_{e\tau}=m_{\mu\tau}$, and
therefore,  $s=c=\displaystyle{\frac{1}{\sqrt{2}}}$.
The $\nu_e-\nu_{\mu}$ level crossing takes place  at $\rho=0$.
As before $\nu_{\tau}-\nu_{2m}$ resonance is also at $\rho\simeq 0$.
The Hamiltonian of $\nu_{\tau}-\nu_{2m}$ system is simplified:
$\nu_{1m}$ decouples since at small densities $\alpha \approx 0$.

Another possibility,  $m_{e\tau}>m_{\mu\tau}$,  is
realized in case of inverse  hierarchy of
couplings, $f_{e\tau}>f_{\mu\tau}$. Now
$\theta>\displaystyle{\frac{\pi}{4}}$ and $\nu_e-\nu_{\mu}$ level crossing
is at negative $\rho$;  $e\mu$-resonance
conversion  takes place in the antineutrino  channel.
Crossing of the $\nu_{\tau}$- $\nu_{2m}$- levels occurs at $\rho\simeq
0$ as before. However
these possibilities are disfavored by data from SN1987A \cite{super}.

\section{Neutrino transitions in medium with variable density.}

{\it Adiabatic conversion.}
Mixing matrix and the eigenvalues obtained in sect.3 allow to get
the probabilities for oscillations in
uniform medium as well as for the adiabatic
conversion in medium with varying density.
The probability of the  adiabatic
conversion $P(\nu_i\rightarrow\nu_j)$ averaged over
oscillations  is determined
by the mixing in the initial moment, $S(\rho_0)$, and in the final moment,
$S(\rho_f)$:
\begin{equation}
P^{ad}(\nu_i\rightarrow\nu_j)=\displaystyle{\sum_k}|S_{jk}(\rho_f)
S_{ik}(\rho_0)|^2 .
\end{equation}

Suppose $\nu_e$ is produced at the density $\rho_0$ and
propagates adiabatically to $\rho = 0$, then
substituting  $S(\rho_0) = S_{12}(\theta_m^0)$  and
$S(\rho_f) = S_0$  in (23) one finds
the $\nu_e \rightarrow \nu_e$  survival probability
\begin{equation}
P^{ad}(\nu_e\rightarrow\nu_e) = (1-\displaystyle{
\frac{3}{2}}s^2)\cos^2 \theta_m^0 + \displaystyle{\frac{1}{2}}s^2,
\end{equation}
where $\theta_m^0 \equiv \theta_m(\rho_0)$ and the mixing angle
$\theta_m$ is determined in (16).
For $\rho_0 \gg 1$ the mixing angle is   $\theta_m^0 \approx
\displaystyle{\frac{\pi}{2}}$,
and from (24) one gets $P \approx \frac{1}{2} s^2$;
it differs by factor
$\displaystyle{\frac{1}{2}}$ from $2\nu$ conversion probability
due to maximal
mixing oscillation (conversion) between two nearly degenerated states.
When $\rho\rightarrow 0$ ($\theta_m\rightarrow\theta$), the probability
(24) reduces to  averaged vacuum probability (11).
Note that at $s^2=\displaystyle{\frac{2}{3}}$ the probability
does not depend on matter effects:   $P^{ad} = 1/3$.
At $s^2<\displaystyle{\frac{2}{3}}$
$\left( s^2>\displaystyle{\frac{2}{3}} \right)$ the probability
$P^{ad}(\nu_e\rightarrow\nu_e)$  decreases (increases)
with $\rho_0$ increase.

The ($\nu_e\rightarrow\nu_{\mu}$)-transition probability,
$P^{ad}(\nu_e\rightarrow\nu_{\mu})$,
can be found from (24)
by the interchange $s^2 \leftrightarrow c^2$;
the interchange $c_m \leftrightarrow s_m$
gives the probability of the
 transition
$P^{ad}(\nu_{\mu}\rightarrow\nu_e)$. The  antineutrino
transitions are also described by (24);
the  mixing angle of antineutrinos
in matter is smaller than that in vacuum,
e.g.  for $\rho_0 \gg 1$ one has
$\theta_m \approx 0$ and
$
P(\bar{\nu}_e \rightarrow \bar{\nu}_e) \approx \cos^2 \theta.
$

\vskip 3mm
\noindent
{\it Adiabaticity violation}.
Let us consider the  general case taking into account
the effects of
adiabaticity violation and  oscillations.
The task is essentially simplified due to the existence
of two different scales of $\Delta m^2$. As it was mentioned in
sect.3  in a given density region
only one mixing angle changes appreciably, whereas two others are
``frozen". Consequently,  the three neutrino task is reduced
to two neutrino tasks. In this case one can introduce partial
adiabaticity parameter $\kappa_{ij}$ that determine the probability of a
jump between  two given levels $H_i$, $H_j$ as
\begin{equation}
\kappa(\rho)_{ij} = \frac{\Delta H_{ij}(\rho)}{\dot{\psi}_{ij}(\rho)}
\cdot \frac{m_0^2}{2 E}~,
\end{equation}
where $\dot{\psi} \equiv \frac{d\psi}{dx}$
determines the change of the level mixing in a given density region
and $\Delta H_{ij}(\rho)$ is the level splitting.

In the region of large densities,
$\rho \gg 4\epsilon/ \tan \theta$,
the change of mixing is stipulated
mainly by $\theta_m$, i.e.
$\psi \equiv \theta_m$,
and in the $e-\mu$ resonant point
one gets using (16):
\begin{equation}
\kappa^R_{12} = \frac{2 \sin^2 2 \theta}{\dot{\rho}}
\cdot \frac{m_0^2}{2 E}~.
\end{equation}
As can be shown for a not too
small mixing angle $\theta$ the adiabaticity for
the 2-3 levels is fulfilled much better than for the 1-2 levels so that
with increasing $\dot{\rho}$ the adiabaticity
starts to be broken first for the 1 - 2 levels and then for the 2 - 3
levels. In the case of complete adiabaticity  for all levels
the neutrino state produced at $\rho_0 \gg 1$
as $\nu_e$ follows the $H_2$
level (fig.1).  If the adiabaticity of $\nu_{2m}^0 - \nu_{\tau}$
system is broken then neutrino evolves along the $H_2^0$ trajectory which is
actually very close to $H_2$ for $\rho \gg \epsilon/s^2$. Consequently,
in this region it does not matter whether the 2 - 3 level
adiabaticity is broken or
not. In case of a strong  adiabaticity violation for 1 - 2 levels in
$e\mu$-crossing region the neutrino state  follows the $\nu_e$-trajectory.

In the region of small densities, $\rho \aprle 4\epsilon/\tan \theta$,
one has $\psi = \beta$.
Here $\theta_m \approx \theta$, $\alpha \approx 0$.
The adiabaticity parameter
for $\nu_{2m}^0 - \nu_{\tau}$ levels
in resonance  ($\rho \approx 0$):
\begin{equation}
\kappa_{23}^R \approx
\frac{32 \epsilon^2 \cos^2 \theta}{\dot{\rho}}\cdot \frac{m_0^2}{2 E}
\end{equation}
is much smaller than $\kappa_{12}(\rho = 0)$, i.e. the
adiabaticity can be broken for  $\nu_{2m}^0 - \nu_{\tau}$ levels,
whereas $\nu_{1m}$ propagates adiabatically.

Another circumstance which simplifies the task is the maximal mixing
of $\nu_{2m}^0$ and $\nu_{\tau}$ levels at $\rho = 0$. This ensures that
the probabilities of the  transitions with zero final density,
$\rho_f = 0$,
averaged over the oscillations do not depend on the
adiabaticity condition  in the $\nu_{2m}^0 - \nu_{\tau}$
system (see Appendix).

\vskip 3mm
\noindent
{\it The probabilities of conversion}.
Keeping in mind possible  applications to the solar, supernova and the
atmospheric  neutrinos we will   consider  the propagation of
the electron neutrino, $\nu_e$, produced at some
density $\rho_0$ towards zero density,
$\rho_f = 0$.
In general the
adiabaticity may be broken in $e\mu$- resonance region
as well as in the
region of small densities ($\rho \sim 0$). Leaving the medium
the neutrinos will  oscillate in vacuum, and moreover, the
oscillations induced by small mass splitting may not be averaged out.

According to (15) the decomposition of the initial neutrino state over the
instantaneous eigenstates  is
\begin{equation}
\nu_0 = \nu_e \approx \cos \theta_m^0 \nu_{1m}^0 + \sin \theta_m^0
\nu_{2m}^0,
\end{equation}
where $\theta_m^0$ is the mixing angle in the production
point (the admixture of $\nu_{3m}$, being of the order of $\epsilon$, is
practically unessential). Let us introduce some density
$\rho'$ ($\rho_b < \rho' \ll 1)$ so that at $\rho < \rho'$ the
1-2 level adiabaticity is restored or the change of the 1-2 mixing is
negligibly small.
As the result of propagation over the large density region one
gets then at $\rho'$ the state
\begin{equation}
\nu(\rho') = (A_{11} c_m^0 + A_{21} s_m^0) e^{i\phi'_m}\nu_{1m}^0(\rho') +
(A_{12} c_m^0 + A_{22} s_m^0) \nu_{2m}^0(\rho'),
\end{equation}
where  $A_{ij}$ (i,j = 1, 2) are the
amplitudes of transitions between the levels
$\nu_{1m}^0$ and $\nu_{2m}^0$, $\phi'_m$
is the phase,  $c_m^0 \equiv \cos\theta^0_m$ etc.. The amplitudes satisfy the
relation:
\begin{equation}
|A_{12}|^2 = |A_{21}|^2 = 1 - |A_{22}|^2 \equiv P_{12}.
\end{equation}
The jump probability $P_{12}$ can be approximated by the Landau-Zener
probability \cite{adiab}  (or its modifications):
\begin{equation}
P_{12} \approx P_{LZ} \equiv exp(-\frac{\pi}{2}\kappa_{12}^R),
\end{equation}
where $\kappa_{12}^R$ is the adiabaticity parameter in resonance (26).
For the adiabatic transitions one has  $A_{ij} = \delta_{ij}$.

In the region  of small densities, $\rho \aprle \rho'$,  the state
$\nu_{1m} \approx \nu_{1m}^0$
propagates adiabatically
so that its admixture does not change, and at zero density
one gets $\nu_{1m} \approx \nu_1$. The evolution of $\nu_{2m}^0$ state is
described by matrix (19). The result of its propagation to zero
density can be presented as
\begin{equation}
\nu_{2m}^0 \rightarrow |a_{22}| e^{i\phi_m} \nu_2
+ |a_{23}| \nu_3 ~,
\end{equation}
where $a_{22}$ and $a_{23}$ are the amplitudes of the transitions
$\nu_{2m} \rightarrow \nu_2$  and $\nu_{2m} \rightarrow \nu_3$
correspondently. In case of the adiabatic propagation
$a_{22} = 1$, $a_{23} = 0$. Note that  neutrino crosses only half of the
2-3 resonance region
and therefore the jump probability
equals half of $P_{LZ}$:
$P_{23} \equiv |a_{23}|^2 \approx \frac{1}{2} P_{LZ}$.
Combining (32) and (29) one gets the  neutrino state
at the edge of  medium:
\begin{equation}
\nu(0) =\left(A_{11} c_m^0 + A_{21} s_m^0\right) e^{i\phi'_m}\nu_{1} +
\left(A_{12} c_m^0 + A_{22} s_m^0\right) \left(|a_{22}| e^{i\phi_m}\nu_2 +
|a_{23}| \nu_3\right).
\end{equation}
Further propagation in vacuum  results in
changes of phases only\footnote{The lost of coherence due to wave packet
spread is reduced to the averaging effect.}:
$\phi_m \rightarrow \phi = \phi_m + \phi_{vac}$ and
$\phi'_m \rightarrow \phi' = \phi'_m + \phi_{vac}'$, where
\begin{equation}
\phi_{vac} = \frac{\Delta m_{23}^2}{2 E} L
\end{equation}
is phase difference acquired at a distance  $L$ in vacuum. Using (33) and
the vacuum
mixing matrix (7) one can obtain the probabilities of different
transitions. In particular, ($\nu_e\rightarrow\nu_e$)-
survival probability averaged over short length oscillations
(phase $\phi'$) is
\begin{equation}
\begin{array}{r}
P(\nu_e \rightarrow \nu_e) \equiv |< \nu_e| \nu(0) >|^2 =
c^2\left(\cos^2 \theta_m^0 - P_{12} \cos 2 \theta_m^0 \right) + \\
+ \frac{1}{2} s^2 \left(\sin^2 \theta_m^0 + P_{12}\cos2\theta_m \right)
\left(1 + 2 |a_{22} a_{23}|\cos \phi \right),
\end{array}
\end{equation}
where we have taken into account (30) as well as
similar normalization condition for $a_{ij}$.
The probability averaged over $\phi$:
\begin{equation}
\bar{P}(\nu_e\rightarrow\nu_e)
= \left(1 - \frac{3}{2}s^2 \right) \left[ \cos^2 \theta_m^{0} -
 P_{12} \cos2\theta_m^0 \right] + \frac{s^2}{2}
\end{equation}
does not depend on $a_{ij}$ at all
in accordance with  the general statement proved in the Appendix.
At $P_{12} = 0$ (adiabatic propagation in the region of large
densities)  the result (36) coincides  with (24).
Using (36) we can rewrite the  probability (35) as
\begin{equation}
P(\nu_e\rightarrow\nu_e)  = \bar{P}
+ s^2 \left(\sin^2 \theta_m^0 + P_{12}\cos2\theta_m^0 \right)
\sqrt{P_{23}(1-P_{23})} \cos \phi,
\end{equation}
where $P_{23} \equiv |a_{23}|^2$. If the adiabaticity for
$\nu_{2m}^0 - \nu_{\tau}$
levels is strongly broken then one has
$a_{22} = a_{23} = \frac{1}{\sqrt{2}}$ or $P_{23} =
\frac{1}{2}$ which corresponds to
$\nu_{2m}^0 \rightarrow \nu_2^0 = \frac{\nu_2  +  \nu_3}{\sqrt{2}}$
transition in medium. Substituting $P_{23} = \frac{1}{2}$ in (37)
one gets
\begin{equation}
P(\nu_e\rightarrow\nu_e) = \bar{P}
+ \frac{1}{2} s^2 \left(\sin^2 \theta_m^0 + P_{12}\cos2\theta_m^0
\right) \cos \phi.
\end{equation}
In this case  the depth of vacuum oscillations is maximal.

\vskip 3mm
\noindent
{\it Energy dependence of the suppression factors.}
Let us consider  the dependence  of the probability (38)
on the  neutrino energy (fig.2).  $P(E)$ is the oscillating
curve inscribed into the band between  $P^{max}$ and $P^{min}$
(the oscillations are stipulated by  change of $\phi$). For
$\cos(\frac{\Delta m^2}{2E}L)=1$ one gets from (38)
\begin{equation}
P^{max}=\displaystyle{\frac{1}{2}+\left(\frac{1}{2}-
P_{12}\right)\cos2\theta_m^0
\cos2\theta}
\end{equation}
which  coincides with 2$\nu$-probability $P_{2\nu}$, i.e.
$P_{2\nu}$ gives the upper bound for $3\nu$-survival probability:
$P_{3\nu} < P_{2\nu}$.
The width of the band,
\begin{equation}
\Delta P \equiv P^{max}-P^{min} = s^2(\sin^2\theta_m^0
+ P_{12}\cos2\theta_m^0),
\end{equation}
is proportional to $\sin^2 \theta$, and consequently, with diminishing
$\theta$ the $3\nu$-probability converges to $2\nu$-probability.
For neutrinos propagating in matter
with monotonously changing density $P^{max} = P_{2\nu}(E)$
has the form of pit (fig.2) \cite{msw}. Outside the pit
the probability approaches 1 on the right hand side and the vacuum value,
$\sin^2 2\theta/2$, on the left hand side. The position of the left
(adiabatic)
edge of the pit, $E_{ad}$, is fixed via the resonant condition
by the density in the production point.
The position of the right (nonadiabatic) edge $E_{na}$ is determined by
the adiabaticity condition.  As follows from (40),
$\Delta P=s^2\sin^2\theta_m^0$ in the adiabatic region.  In
particular, for  small energies outside the pit, where the matter effect
is weak, one has $\Delta P=s^4$. Maximal width of the strip,
$\Delta P=s^2$, is at the bottom of the pit when $\rho_0>>\rho_R$,
therefore at the bottom: $P^{min}=0$.
 In the nonadiabatic region
$\Delta P$ decreases with enhancement of the adiabaticity
violation: $\Delta P\simeq s^2(1-P_{12})$.

The position of the first (broadest)
minimum of the oscillating curve, $E_m$, is determined from the condition
$\displaystyle{\frac{l_{\nu}}{2} = L}$, where $l_{\nu}$ is the
oscillation length in vacuum.
Explicitly
one has  $\displaystyle{{E_m} = \frac{\Delta m^2_{23} L}{2 \pi}}$;
the first maximum is at $E_m/2$ etc..
Mutual position of the pit and
the modulating curve is fixed by  the ratio of energies
\begin{equation}
\frac{E_{ad}}{E_m} =
\frac{2 \pi \cos 2\theta}{L \sqrt{2} G_F n_0}
\frac{\Delta m^2_{12}}{\Delta m^2_{23}}
\approx
\frac{\Delta m^2_{12} l_0}{\Delta m^2_{23} L },
\end{equation}
where $l_0 \equiv 2\pi /\sqrt{2} G_F n_0$ is the refraction length
in the neutrino production point.
If $E_m \gg E_{na} > E_{ad}$, fast oscillations of the modulating curve
are practically averaged in the
energy region of the pit and $P \approx \bar{P}$. For $E_m \sim E_{ad}$
one predicts  the observable  modulations of the pit.
If $E_{m} \ll E_a$ long length
oscillations are not developed and $P \rightarrow P_{2\nu}$.

Similarly one can analyze the transitions of
$\nu_{\mu}$ and $\nu_{\tau}$. In particular,
$\nu_{\tau}$ is converted mainly in the region of small
densities,  and the probability of $\nu_{\tau} \rightarrow \nu_e$
transitions averaged over large scale splitting equals
\begin{equation}
P(\nu_{\tau} \rightarrow \nu_e)
 = \frac{s^2}{2}\left[1 + \sqrt{P_{23}(1-P_{23})} \cos \phi \right].
\end{equation}
For the antineutrino channel (negative $\rho$) there is no
$\nu_e - \nu_{\mu}$ level crossing; one can consider
adiabatic evolution of
$\bar{\nu}_{1m}$ and  $\bar{\nu}_{2m}^0$ states and as the result
$P(\bar{\nu}_e \rightarrow \bar{\nu}_e)
\approx P^{ad} = \cos^2 \theta$.

If $f_{e\tau} \sim f_{\mu \tau}$, the $e \mu$-mixing
becomes maximal; both resonant regions are at $\rho = 0$.
The above consideration (reduction to two neutrino tasks)
is valid due to difference in the widths of the resonance layers.
Indeed in $\nu_{\tau} - \nu_{2m}^0$ crossing region  the change of the
$\nu_e - \nu_{\mu}$ mixing is negligibly small.

\section{Applications}

The results of the solar  neutrino experiments can be reconciled with
predictions of the Standard Solar Models
in terms of the resonant flavor conversion
$\nu_e\rightarrow\nu_{\mu}(\nu_{\tau})$
with parameters (two neutrino mixing):
$\Delta m^2=(0.4-1.2)\cdot 10^{-5} {\rm eV}^2$,
$\sin^22\theta=(0.1-1.5)\cdot 10^{-2}$
``small mixing solution" and
$\Delta m^2=(0.5 - 3)\cdot 10^{-5} {\rm eV}^2$,
$\sin^22\theta=(0.60-0.85)$
``large mixing solution" \cite{sun}.
However taking into account the uncertainties of both the predictions and
the experimental data, one should consider more wide region of the
parameters.
The deficit of $\nu_{\mu}$ in the atmospheric neutrino flux testifies for
oscillations $\nu_{\mu}\rightarrow \nu_{\tau}$, with
$\Delta m^2=(0.5-3)\cdot 10^{-2} {\rm eV}^2$,
$\sin^22\theta=(0.4-0.6)$
or for oscillations $\nu_{\mu}\rightarrow \nu_{e}$ with
$\Delta m^2=(0.5-3)\cdot 10^{-2} {\rm eV}^2$,
$\sin^22\theta=(0.3-0.8)$ \cite{atm}.
The formation of large scale structure of the Universe implies
the existence of hot component of dark matter and the relic neutrinos
with $m\sim (2-7)$ eV can play such a role \cite{rel}.

Let us confront these results with predictions of the Zee-model for
different values of $m_0$.
As follows from (2, 4), basic mass scale is
$
m_0 = 7 \cdot 10^4 \cdot \left(I \sqrt{f_{e\tau}^2  + f_{\mu\tau}^2}
\right)
$ eV,  therefore
depending on values of couplings $f_{ij}$ as well as the  parameters of the
scalar sector one can get for $m_0$  any value below $\sim 10^4$ eV.
For $f_{e\mu} \leq f_{e\tau} \leq f_{\mu \tau}$
the ratio of the mass squared difference (8)
may be in the range $10^{-5} - 5 \cdot 10^{-3}$,
unless one introduces very strong hierarchy of couplings.
Also under this condition the mixing angle is in the region
$\sin^2 2\theta = 10^{-3} - 1$.

Three regions of $m_0$ values are of special interest.

\vskip 3mm
\noindent
{\it Cosmologically interesting mass scale}: $m_0 \sim (1 - 30)$ eV.
The components $\nu_2$ and $\nu_3$  with masses $m_0$,
can form  the hot  dark matter. Since $\Delta m_{13}^2\sim
\Delta m_{12}^2 \sim m_0^2\aprge 1 \rm{eV}^2$, the  mixing angle,
$\theta$, is restricted
by the   accelerator oscillation  experiments: e.g.
$\sin^22\theta<2\cdot 10^{-3}$ at $m_0=3$ eV,  $\sin^22\theta<6\cdot 10^
{-3}$ at $m_0=1$ eV  etc. \cite{osc}.
For $\sin^2 2\theta > 10^{-5}$
one expects strong resonant conversions
$\nu_e \rightarrow \nu_{\mu}$
and $\nu_{\mu} \rightarrow \nu_e$  in the
inner parts of the collapsing star. Such a conversion results in
permutation of the  $\nu_e$- and $\nu_{\mu}$- energy spectra and
therefore in the increase of average energy of the electron neutrinos.
This will have two consequences:
(i) the increase of the energy release due $\nu_e
- e$- scattering which may help to expel the envelope, (ii) the
formation
of the proton-rich medium due to dominant $\nu_e n \rightarrow e p$
scattering. The latter will forbid the r-processes responsible for
nucleosynthesis of heavy elements \cite{qian}.
If the inner part of collapsing stars is the only place of the
r-processes, then the indicated conversion should be suppressed and
one gets the bound on mixing angle
$\sin^2 2\theta < 10^{-4} - 10^{-5}$ \cite{qian}.
The value of  $\Delta m_{23}^2$, can be naturally in the region
responsible for the atmospheric neutrino problem.
The suppression of the muon neutrino flux
due to  $\nu_{\mu} - \nu_{\tau}$-oscillations
is determined by the averaged vacuum probability
(11) and  taking into
account the indicated bounds on $\theta$ one gets:
$P\approx 0.5$ which is actually  outside the region of the best
fit of all the data.
The suppression of the solar $\nu_e$-flux fixed by averaged
vacuum probability (11) is very weak:
 $P(\nu_e\rightarrow\nu_e) \geq 0.92$.

\vskip 3mm
\noindent
{\it Atmospheric neutrino mass scale}:
$m_0\simeq (0.3-1)\cdot 10^{-1}$eV.
 $m_0^2$ is in the region of the solution of the atmospheric
neutrino problem $\Delta m_{13}^2 \approx m_0^2
\simeq (10^{-3}-10^{-2})$eV$^2$.
The deficit of $\nu_{\mu}$ can be
explained by $\nu_{\mu} \leftrightarrow \nu_e$ - oscillations, with
$\sin^22\theta=0.3 - 0.8$.

For  $\Delta m^2_{23} =  (10^{-6} - 10^{-4})$ eV$^2$
the suppression factor for solar neutrinos is
determined by the adiabatic probability (24)
which coincides  at $\theta_m \approx \theta$
with averaged vacuum
probability  (11).  Solar neutrino spectrum is outside the pit at
small energies. Although $\Delta m_{23}^2$ is in
the region of strong  matter effect, the averaged probability
$P_{\odot}$, practically does not depend on matter density
and on the neutrino energy as well.
For values of $\theta$ needed to solve
the atmospheric neutrino problem (at $s^2 < c^2$) one gets from (11)
$P_{\odot}=0.56 - 0.83$.
Taking into account an additional contribution to the
$\nu e$-scattering from neutral currents one  predicts the following
suppression factors $(R \equiv data/SSM)$ for
Ga- , Ar-, production rates and $\nu e$-signal:
\begin{equation}
R_{Ge}=R_{Ar}\simeq P_{\odot}\sim 0.56-0.83, ~~~~
R_{\nu e}\simeq 0.62 - 0.86.
\end{equation}
There is no distortion of  energy spectrum  of boron neutrinos.
The predictions (43) fit rather well all the results except the one of
Homestake experiment.

In case of very small splitting $\Delta m_{23}^2 \ll 10^{-9}$ eV$^2$, there
is no averaging over long length oscillations and
the  probability is modulated with the amplitude $s^4/2$
 (see (40) and further discussion).
For the indicated mixing angles the amplitude of modulations is  $\sim (0.25 -
4.5) \cdot 10^{-2}$. So, one may expect up to
$\sim 10\%$ distortion of boron neutrino spectrum and change of the
relations (43). For example, at $\sin^2 2\theta = 0.8$
and for certain values of $\Delta m^2$  one may get
$R_{Ge}= 0.54$,  $R_{Ar} =  0.51$,  and $R_{\nu_e} =  0.58$.
Further increase of $\sin^2 2\theta$,
although enhances the amplitude of modulations, results in
the stronger suppression of the atmospheric $\nu_{\mu}$-signal as well as
the signal in gallium experiments.

As in the previous case one predicts strong resonant conversions
$\nu_e \rightarrow \nu_{\mu}$
and $\nu_{\mu} \rightarrow \nu_e$ of the neutrinos from the collapsing
cores of stars. The corresponding permutation factor which characterizes
the interchange of $\nu_e$- and $\nu_{\mu}$- energy spectra \cite{super}
equals $p = 0.75 - 0.90$. Moreover, since the mixing is rather large
one expects an appreciable permutation  of the antineutrino spectra:
$\bar{\nu}_e \leftrightarrow \bar{\nu}_{\mu}$; the permutation factor
$\bar{p} = 0.08 - 0.25$ results in even better description of data
from SN1987A \cite{permut}, \cite{super}.

Note that for indicated values of $m_0$ the conversions take place now in more
external layers,  so that there is no problem with r-processes.

Another value of the mixing angle which follows from explanation of
the atmospheric neutrino data ($s^2 > c^2$, this
corresponds to the consideration in \cite{kras})
gives $P_{\odot} = 0.17 - 0.44$ and $P_{\nu e} = 0.29 - 0.51$.
The result contradicts to the  observed gallium
production rate. Moreover  the $e-\mu$-resonance is in the
antineutrino channel which results in strong
permutation (
$\bar{p} = 0.6 - 0.8$) of the
$\nu_e-, \nu_{\mu}-$ energy spectra. The SN1987A data
give the bound~ $\bar{p} < 0.4$ \cite{super}.

\vskip 3mm
\noindent
{\it Solar neutrinos mass scale}:
$m_0\simeq (10^{-2}\sim 10^{-3})$ eV.  the largest mass splitting,
$\Delta m_{13}^2 \sim (10^{-6} - 10^{-4})$ eV$^2$,
is in the region of the resonant effect inside the Sun.  If
$\Delta m_{23}^2>>10^{-10}$eV$^2$,  averaging over long length
vacuum oscillations  takes place and the suppression factor for
$\nu_e$-flux is determined by the probability  $\bar{P}$ (36). It can be
rewritten as:
\begin{equation}
P= P_{2\nu} - \frac{s^2}{2}\left[\sin^2 \theta_m  + P_{12} \cos 2\theta_m
\right], \end{equation}
where $P_{2\nu}$ is the suppression factor for two neutrino mixing.
One can easily construct the $3\nu$-suppression pit
using  (44) and the results for two neutrino mixing.
 For small $\theta$ the effect of the third neutrino is negligibly small and
$P\simeq P_{2\nu}$. The Zee-mass matrix reproduces small mixing
solution of the solar neutrino problem for two neutrinos.
The deviation from the 2$\nu$ case
is of the order of  $10^{-3}$ and to distinguish the Zee mechanism from the
other mechanisms one can take into account
the following facts. In the considered case there
is no oscillation solution of the atmospheric neutrino problem as well as
no appreciable contribution of
neutrinos to hot dark matter. Moreover, there is no manifestation
of the third neutrino in the experiments with supernova neutrinos.

Large mixing solution is absent unless one admits large original flux
of boron neutrinos. Indeed, since at the bottom of the pit $P =
\sin^2 \theta /2$ one needs two times bigger
value of $\sin^2 \theta$ in comparison with 2$\nu$ case
to get the same suppression of Kamiokande  signal, e.g.  instead  of
$\sin^2 2\theta = 0.7$ one should take
$\sin^2 2\theta = 0.994$. However  in this case the fluxes  of
the low energy neutrinos are strongly suppressed:
in the pp-neutrino region one gets $P \approx 0.375$.

The situation is essentially different if
$\Delta m_{23}^2\leq 10^{-9}$eV$^2$, and consequently
there is no averaging over the long-length oscillation at
least in some part of the suppression pit (fig.2).
One of the most interesting
configurations is shown in fig.3 which corresponds to
$\Delta m_{23}^2/\Delta m_{12}^2 \sim  10^{-5}$ and
$\sin^2 2\theta = 0.8 - 0.9$.  First minimum of
oscillating curve is at adiabatic edge, first maximum is
outside the pit. If the  Be-neutrino line is  in
first minimum of oscillating curve
then the boron neutrinos are at  the bottom of
suppression pit outside the minimum,
and the detected part of the
pp- neutrino spectrum is in the
first maximum. The signature of such a scenario
is the strong suppression of the Be-neutrino flux, and
the absence of the  {\it distortion} of the high energy part of the boron
neutrino spectrum in contrast
with small mixing solution for two neutrinos.
For pp-neutrinos one can get the suppression 0.55 -
0.60, so that total Ge-production rate could be about 50 - 70 SNU in
agreement with present data.

Another possibility corresponds to  the modulating  curve  shifted to
larger energies, so that boron neutrinos are in the first maximum
(pp-neutrinos are in averaging region) and the beryllium line is
in the fastly
oscillating region.
If the oscillating curve is shifted to lower energies then one
expect the distortion of the pp-neutrino spectrum.

Obviously in this case there is no solution of the atmospheric neutrino
problem.
For the indicated values of parameters one may expect complete or partial
conversion of the supernova neutrinos $\nu_e \leftrightarrow \nu_{\mu}$
depending on  density profile of  star.

The loop diagrams, similar to those generating the neutrino masses,
will  generate also the transition magnetic moments of
the neutrinos. These moments are  however restricted by $\mu < 3 \cdot
10^{-15}
\mu_B (m_0/1 \rm{eV})$, where $\mu_B$ is Bohr magneton \cite{mag}, so that
even for $m_0 \approx 10$ eV the effects of  spin-flip on
the solar and atmospheric neutrinos are
negligibly small.

\vskip 3mm
\noindent
{\it Implications to the parameters of the Zee model}.
Let us comment on possible implications of the above results to the
original Zee-model. The expression for $m_0$ can be rewritten as
\begin{equation}
f_{\mu\tau} \sim 1.3 \cdot 10^{-5} \left(\frac{m_0}
{1 \rm{eV}}\right)\cdot
\frac{1}{I}~.
\end{equation}
Moreover, the dimensionless parameter $I$ can be of the order  1 when
all mass parameters entering $I$ are at the electroweak scale and
$v_u \sim v_d$. According to (45) a value $I = 1$ gives the lower bound
on $f_{\mu\tau}$. In principle, $f_{\alpha\beta}$ can be as large as 1;
$f_{\mu\tau} \approx 1$ gives the lower bound on $I$, and consequently,
the upper bound on the mass of charged scalar
$M_2$. Let us estimate the range of the parameters.

For $m_0$ in cosmologically interesting domain one has
$f_{\mu \tau} >  (0.13 - 4)\cdot 10^{-4}$ and
$M_2 < (2 - 8)\cdot10^4$ GeV.
The indicated values  of mass squared differences  and
accelerator bounds on $\theta$ correspond to
$f_{e\mu} \sim f_{e \tau} \sim (0.03 - 0.1) f_{\mu \tau}$.
If $m_0$ is in the region of the atmospheric neutrino problem
one gets $f_{\mu \tau} > (0.4 - 1.3) \cdot 10^{-6}$
and  $M_2 < (0.8 - 2.4)\cdot10^5$ GeV.
All constants can be of the same order:
$f_{\mu \tau} \sim  f_{e\tau} \sim  f_{e \mu}$.
Alternatively, if there is
no averaging over long length oscillations
$f_{e \mu} = 10^{-4}f_{\mu \tau}$.
For $m_0$ in the region of the solutions of the solar neutrino problem
the parameters are $f_{\mu \tau} > (0.13 - 1.3)\cdot 10^{-8}$
and $M_2 < (3 - 4)\cdot10^6$ GeV. Small mixing solution
implies $f_{e\tau} < 0.1 f_{\mu \tau}$. Large mixing solution
with modulations by long length oscillations  is realized when
$f_{e\tau} \sim  f_{\mu \tau}$ and $f_{e \mu} \sim 10^{-2}f_{\mu \tau}$.

\section{Discussion  and Conclusions}

1. Zee-matrix in matter (for $\epsilon \ll 1$) is  an example of
``solvable" $3\nu$-task.
This allows to trace some interesting features of the dynamic of
propagation, in particular, the effect of strong influence of one
resonance on another. Dominant $\nu_e - \nu_{\mu}$ mixing shifts the
resonance for  $\nu_{\tau}$ to zero density, thus changing a naive
picture of  level crossings.

2. In the supersymmetric generalization of the model
new diagrams appear with sleptons and higgsino
in the loops. General structure of mass matrix is the same as (1).
Note that Zee singlet can be embedded in the SU(5) GUT
scheme by introducing the antisymmetric 10-plet of scalars.

3. Zee-matrix  can be considered as an alternative to the one
generated by the see-saw mechanism. Let us note for a sake of completeness
that in principle the matrix (1) can be reproduced by the see-saw
mechanism too. For this
one needs a special structure of the Majorana mass matrix of the right
components, $M_R$. Namely, in the Dirac neutrino basis
(where the  Dirac matrix is diagonal) $M_R$ should have zero determinants
of three submatrices: $M_{ii} M_{jj} - M_{ij}^2 = 0$, (i,j = 1,2,3).

4. The mass matrices generated at the  one loop level in the model with
explicit R-parity violation \cite{rpar} also differ from (1).
The generic property of these matrices  is the existence of
nonzero diagonal elements which are not suppressed in comparison with
nondiagonal elements.
If purely lepton couplings dominate over couplings of quark and lepton
supermultiplets, the elements $m_{e\tau}$,
$m_{\mu\tau}$,
are naturally suppressed with respect to others by factor
of $m_{\mu}/m_{\tau}$. Moreover, the element $m_{\tau\tau}$
has even  stronger suppression: $(m_{\mu}/m_{\tau})^2$.
Such a  matrix allows to
explain both  solar and the atmospheric neutrino deficits
\cite{smir} according to the scenario suggested in \cite{kras} .

Similar structure of mass matrix appears in the model with
two loops generation of the neutrino masses  \cite{babu}.

5. Practically all  the extensions of the standard model imply the
existence of  neutral fermions which can play the role of
 right-handed  neutrinos. In this case the see-saw mechanism
is obtained and in addition to the
radiative mass terms one gets the  see-saw
contributions to the neutrino mass matrix:
$
m_{ss} \approx m_D M^{-1} m_D^T.
$
The biggest term is $m_{\tau \tau} \approx \frac{m_{3D}^2}{M}$,
where the Dirac mass, $m_{3D}$,  can be  as large  as the top quark mass.
Depending on values of parameters which may be in the range
$m_D = (1 - 10^2)$ GeV and $M = (10^{10} - 10^{18})$ GeV one can get
negligibly small, or comparable with $\epsilon m_0$, or even
dominant  see-saw contribution.
For example, at  $m_{3D} \sim 100$ GeV
and $M =  10^{18}$ GeV one has
$m_{\tau \tau} \sim 10^{-5}$ eV which is of the order of
$\epsilon m_0$ for $m_0 < 10^{-3}$ eV. If  $M <  10^{16}$ GeV
this contribution becomes dominant.

Let us comment on the simplest possibility when only one additional
element, $m_{\tau \tau}$, is important. Such a  situation is realized
if  the Majorana mass matrix of the RH-components
has no strong hierarchy.
In this case one can get easily  both
large $\mu - \tau$ - mixing angle, $\theta_{\mu\tau}$,
needed to solve the atmospheric
neutrino problem and small mixing solution of the solar neutrino
problem.  Indeed, now  $\mu \tau$-mixing is of the order of  $m_0/m_{ss}$
and  large value of $\theta_{\mu\tau}$ follows from the fact that the
see-saw and the radiative contributions are of the same order.
Moreover, for small $e\mu$ mixing the
following relation between the ratio of masses  and the
$\mu \tau$- mixing exists:
\begin{equation}
\frac{\Delta m^2_{\odot}}{\Delta m^2_{atm}} \sim
\frac{m^2_2}{m^2_3} \sim
\left[\frac
{1 - \cos 2\theta_{\mu\tau}}
{1 + \cos 2\theta_{\mu\tau}} \right]^2 .
\end{equation}
For $\sin^2 2\theta_{\mu \tau} > 0.4$ one gets
$\frac{m^2_2}{m^2_3} > 0.01$ which is roughly consistent with
desirable value.
The ratio can be further corrected if one takes into account
the see-saw contributions to other elements of matrix.

\vskip 3mm
\noindent
{\it In conclusion}, we have considered the
properties of propagation of the neutrinos with
Zee-like mass matrix. Crucial features of the matrix  are zero
(small) diagonal
elements and natural hierarchy of nondiagonal elements.
It  can be considered as an alternative to
the matrix generated
by the see-saw mechanism as well to other radiative mechanisms.
The probabilities of the conversions induced by the Zee-mass matrix
as the functions of the neutrino energies can be
represented as two neutrino probabilities
modulated by the oscillating curve related to the long length
oscillations.

The Zee-mass  matrix does  not allow to explain all three problems
related to the solar atmospheric and relic neutrinos simultaneously.
If two heavy components are in the cosmologically interesting domain
then one can get for atmospheric muon neutrinos the suppression 1/2
with no appreciable effect for solar neutrinos.
The Zee-mass matrix allows to  fit well the atmospheric neutrino
data  if the masses of heavy components are in the region $\sim 0.1$ eV.
In this case one predicts energy independent suppression
of $\nu_e$ flux at the level 0.6. For smaller $m_0$ the
matrix can reproduce with high precision small 2$\nu$-mixing
solution  of the solar neutrino problem without any appreciable
manifestations of the third neutrino.
For large mixing $\theta$ new configurations of the
suppression appear. In particular, one may have strong suppression of the
beryllium line and energy independent suppression of the high energy part
of boron neutrino spectrum. In general, the modulations of the smooth
energy dependence of the probabilities for $2\nu$-case  are expected.

These features may allow to identify the Zee-mass matrix
and consequently the corresponding mechanism of neutrino mass generation.

\noindent{\Large \bf Acknowledgement}

The authors would like to thank J. Peltoniemi, S. T. Petcov and  G.
Senjanovi\'c   for  valuable discussions.

\noindent{\Large \bf Appendix}

Suppose the neutrino propagates in medium
with density decreasing from $\rho_i$ in the  initial point
to zero. Let us show that  the averaged
oscillation probability does not depend on the adiabaticity condition for
the levels having maximal mixing in vacuum.

Indeed, the initial neutrino state can be written as
$$
\nu_i = a \nu_{1m} + b \nu_{2m} + c \nu_{3m}.
$$
Suppose $\nu_{1m}$- state propagates adiabatically, whereas
the adiabaticity  for $\nu_{2m} - \nu_{3m}$
may be broken. The latter means that
there are the  transitions $\nu_{2m} \leftrightarrow
\nu_{3m}$ while the neutrino propagates to zero density.
At zero density the neutrino state will have the form
$$
\nu_f = a \nu_{1} + b' \nu_{2} + c' \nu_{3}.
$$
In general $|b'|$ and $|c'|$  differ from $|b|$ and $|c|$,
however  the normalization condition  implies that
$$
|b|^2 + |c|^2 = |b'|^2 + |c'|^2 = 1 - |a|^2.
$$
The probability to find the neutrino
$\nu_\alpha\equiv a_\alpha \nu_{1} + b_\alpha\nu_{2} + c_\alpha\nu_{3}$
in final state averaged over the oscillations equals
$$
P = |a_\alpha^{\dagger}
a|^2 + |b_\alpha^{\dagger}
b'|^2 + |c_\alpha^{\dagger} c'|^2.
$$
Maximal mixing
of $\nu_2$ and $\nu_3$ in $\nu_f$ means that $b_\alpha = c_\alpha$,
and then
taking into account the  normalization condition one gets
$$
P = |a_\alpha^{\dagger}a|^2 + |b_\alpha|^2(|b|^2 + |c|^2) =
|a_\alpha^{\dagger}a|^2 + |b_\alpha|^2(1 - |a|^2)
$$
which  does not depend on
changes of $b$ and $c$.

\newpage
\centerline{ \Large \bf Figure
Caption} \vskip 1cm

Fig.1 \hskip 3mm  Energy levels of the neutrinos with Zee-mass matrix
as the functions of the matter density $\rho$ (full lines).
Dashed lines correspond to the levels in zero approximation over
$\epsilon$. Dotted lines show the flavor levels.
a). Small mixing angle $\theta$,
b). for $\theta = 45^0$.

Fig.2 \hskip 3mm  Survival probability $P(\nu_e \rightarrow \nu_e)$
as the function of the neutrino energy for different values of
mixing angle $\theta$ (solid lines) a).
$\sin^2 2\theta = 0.64$,
b).$\sin^2 2\theta = 0.84$.
Dashed lines correspond to $P^{max}$ and $P^{min}$, the averaged
probability is shown by dotted line.

Fig.3 \hskip 3mm
The suppression factor for
large mixing and nonaveraged vacuum oscillations.
Also shown is  the solar neutrino spectrum (hatched).

\end{document}